\documentclass[conference]{IEEEtran}
\usepackage{blindtext, graphicx}
\usepackage{cite}
\usepackage[cmex10]{amsmath}
\usepackage{amsfonts,amsthm,amssymb}
\usepackage{algorithmic}
\usepackage{caption}
\usepackage[font=footnotesize]{subfig}
\usepackage{graphicx,float,wrapfig}
\usepackage{graphicx}
\usepackage{graphics,color}
\usepackage{mathtools}
\usepackage{nicefrac}
\usepackage{bm}
\usepackage{txfonts}
\usepackage{wrapfig} 
\usepackage{pgfplots}
\usepackage{mathpazo} 
\usepackage[T1]{fontenc}
\usepackage{upquote}
\usepackage[none]{hyphenat}

\DeclareMathOperator{\Tr}{tr}

\newtheorem{theorem}{Theorem}[section]
\newtheorem{definition}[theorem]{Definition}

\numberwithin{equation}{section}

\begin{document}
%
\title{Super-resolution Line Spectrum Estimation \\with Block Priors}

\author{\IEEEauthorblockN{Kumar Vijay Mishra, Myung Cho, Anton Kruger and Weiyu Xu}
\IEEEauthorblockA{Department of Electrical and Computer Engineering\\
The University of Iowa, Iowa City, USA}
}

\maketitle

\begin{abstract}
We address the problem of super-resolution line spectrum estimation of an undersampled signal with block prior information. The component frequencies of the signal are assumed to take arbitrary continuous values in known frequency blocks. We formulate a general semidefinite program to recover these continuous-valued frequencies using theories of positive trigonometric polynomials. The proposed semidefinite program achieves super-resolution frequency recovery by taking advantage of known structures of frequency blocks. Numerical experiments show great performance enhancements using our method.
\end{abstract}


\section{Introduction}
\label{sec:intro}
In many areas of engineering and science, it is desired to infer the spectral contents of a measured signal. In the absence of any \textit{a priori} knowledge of the underlying statistics or structure of the signal, spectral estimation is a subjective craft \cite{marple1987digital}\cite{stoica2005spectral}. However, in several applications, the knowledge of signal characteristics is available through previous measurements or prior research. By including such prior knowledge during spectrum estimation process, it is possible to mitigate the subjective character of spectral analysis.

One useful signal attribute could be its sparsity in spectral domain. In recent years, spectral estimation methods that harness the spectral sparsity of signals have attracted considerable interest \cite{mishali2010theory}\cite{duarte2013spectral}\cite{candes2013towards}\cite{tang2012csotg}. These methods trace their origins to \textit{compressed sensing} that allows accurate recovery of signals sampled at sub-Nyquist rate \cite{donoho2006compressed}. In the particular context of spectral estimation, the signal is assumed to have sparsity in a finite discrete dictionary such as Discrete Fourier Transform (DFT). As long as the true signal frequency lies in the center of a DFT bin, the discretization in frequency domain faithfully represents the continuous reality of the true measurement. If the true frequency is not located on this discrete frequency grid, then the aforementioned assumption of sparsity in DFT domain is no longer valid \cite{tan2014sparse} \cite{huang2012adaptive}. The result is an approximation error in spectral estimation often referred to as scalloping loss \cite{harris1978use}, basis mismatch \cite{chi2011sensitivity}, and gridding error \cite{fannjiang2012coherence}.

Recent state-of-the-art research \cite{candes2013towards}\cite{tang2012csotg}\cite{tang2013justdiscretize} has addressed the problem of basis mismatch by proposing compressed sensing in continuous spectral domain. This \textit{off-the-grid compressed sensing} approach \cite{candes2013towards}\cite{tang2012csotg} uses atomic norm minimization to recover super-resolution frequencies - lying anywhere in the continuous domain $[0, 1]$ - with few random time samples of the spectrally sparse signal, provided the line spectrum maintains a nominal separation.

However, this formulation of off-the-grid compressed sensing assumes no prior knowledge of signal other than sparsity in spectrum. In fact, in many applications, where signal frequencies lie in continuous domain such as radar \cite{skolnik2008radar}, acoustics \cite{trivett1981modified}, communications \cite{beygi2014multiscale}, and power systems \cite{zygarlicki2012prony}, additional prior information of signal spectrum might be available. Of particular interest to spectral estimation are spectrally block sparse signals where certain frequency bands are known to contain all the spectral contents of the signal. For example, a radar engineer might know the characteristic speed with which a fighter aircraft flies. This knowledge then places the engineer in a position to point out the ballpark location of the echo from the aircraft in the Doppler frequency spectrum. Similarly, in a precipitation radar, the spectrum widths of echoes from certain weather phenomena (tornadoes or severe storms) are known from previous observations \cite{doviak1993doppler}. This raises the question whether we can use signal structures beyond sparsity to improve the performance of spectrum estimation.

There are extensive works in compressed sensing literature which discuss recovering sparse signals using secondary signal support structures, such as \textit{structured sparsity} \cite{cevher2009recovery} (tree-sparsity \cite{baraniuk2010model}, block sparsity \cite{stojnic2009reconstruction}, and Ising models \cite{cevher2008sparse}), spike trains \cite{hegde2009compressive} \cite{azais2013spike}, nonuniform sparsity \cite{amin2009weighted} \cite{vaswani2010modified},  and multiple measurement vectors (MMVs) \cite{duarte2011structured}. However, for spectrum estimation, frequency parameters take continuous values rather than discrete support as in compressed sensing. Therefore, the techniques of using prior support information in discrete compressed sensing do not directly extend to spectrum estimation. Moreover, it is rather unclear as to how general signal structure constraints can be imposed for super-resolution recovery of continuous-valued frequency components.

In this paper, we propose a general way to perform super-resolution spectrum estimation, given the prior information about frequency bands within which true frequency components reside. We propose a precise semidefinite program for the atomic norm minimization to recover the frequency components. The key is to transform the dual of atomic norm minimization to a semidefinite program through linear matrix inequalities (LMI). These linear matrix inequalities are in turn provided by theories of positive trigonometric polynomials \cite{fejer1915uber}. Our new method has shown great performance improvement compared with methods using only signal sparsity information.

\subsection{Relationship with prior work}
\label{subsec:priorwork}
There are a number of existing approaches for spectral estimation by including known signal characteristics in the estimation process. The classical Prony's method can be easily modified to account for known frequencies \cite{trivett1981modified}. Variants of the subspace-based frequency estimation methods such as MUSIC and ESPRIT have also been formulated \cite{linebarger1995incorporating} \cite{wirfalt2011subspace}, where prior knowledge can be incorporated for parameter estimation. For applications wherein only approximate knowledge of the frequencies is available, the spectral estimation described in \cite{zachariah2013line} applies circular von Mises probability distribution on the spectrum. For irregularly spaced samples, sparse signal recovery methods which leverage on prior information have recently gained attention for general applications \cite{amin2009weighted} \cite{vaswani2010modified} \cite{ji2008bayesian} as well as, specifically, spectral estimation \cite{bourguignon2007sparsity}. Compressed sensing with clustered priors was addressed in \cite{yu2012bayesian} where the prior information on the number of clusters and the size of each cluster was assumed to be unknown.

Also, a number of generalizations of off-the-grid compressed sensing for specific signal scenarios have been attempted. A generalization to two-dimensional off-the-grid frequencies involves Hankel matrix completion, and guarantees robustness against corruption of data \cite{chi2013robust}. A heuristic approximation to atomic norm minimization for two-dimensional off-the-grid frequencies was proposed in \cite{chi2013compressive}. Later, the open problem of precise optimization formulation of atomic norm minimization for two and higher dimensions was solved in \cite{xu2013precise} using theories of multivariate positive trigonometric polynomials. When some frequencies are \emph{precisely} known, \cite{mishra2013off} proposed to use \textit{conditional atomic norm} minimization to recover the off-the-grid frequencies. However, in practice, frequency components are seldom precisely known, and more often frequency locations are only approximately known. In this paper, we greatly broaden the scope of prior information by considering \textit{block priors} -frequency subbands in which true spectral contents of the signal are known to exist - to enhance spectrum estimation performance.

\section{System Model}
\label{sec:sys_mod}
We consider a frequency-sparse signal $x[l]$ expressed as a sum of $s$ complex exponentials,
\begin{equation}
\label{eq:sigmodelstd}
x[l] = \sum\limits_{j=1}^{s} c_je^{i2\pi f_jl} = \sum\limits_{j=1}^{s} |c_j|a(f_j, \phi_j)[l]\phantom{1}, \phantom{1} l \in \mathcal{N}
\end{equation}
where $c_j = |c_j|e^{i\phi_j}$ ($i = \sqrt{-1}$) represents the complex coefficient of the frequency $f_j \in [0, 1]$, with amplitude $|c_j| > 0$, phase $\phi_j \in [0, 2\pi)$, and frequency-\textit{atom} $a(f_j, \phi_j)[l] = e^{i(2\pi f_j l + \phi_j)}$. We use the index set $\mathcal{N} = \{l\phantom{1}|\phantom{1} 0 \le l \le n-1\}$, where $|\mathcal{N}| = n, n \in \mathbb{N}$, to represent the time samples of the signal.

We further suppose that the signal in (\ref{eq:sigmodelstd}) is observed on the index set $\mathcal{M} \subset \mathcal{N}$, $|\mathcal{M}| = m \ll n$ where $m$ observations are chosen uniformly at random. Then, the off-the-grid compressed sensing problem is to recover all the continuous frequencies with very high accuracy using this undersampled signal.

\subsection{Off-the-grid compressed sensing}
\label{subsec:otgcs}
The signal in (\ref{eq:sigmodelstd}) can be modeled as a positive linear combination of the unit-norm frequency-\textit{atoms} $a(f_j, \phi_j)[l] \in \mathcal{A} \subset \mathbb{C}^n$ where $\mathcal{A}$ is the set of all frequency-atoms. These frequency atoms are basic units for synthesizing the frequency-sparse signal. This definition of frequency-sparse signal leads to the following formulation of the \textit{atomic norm} $||\hat{x}||_\mathcal{A}$ - a sparsity-enforcing analog of $\ell_1$ norm for a general atomic set $\mathcal{A}$:
\begin{equation}
\label{eq:atomicnorm}
||\hat{x}||_{\mathcal{A}} = \underset{c_j, f_j}{\text{inf}}\phantom{1}\left\{\sum\limits_{j=1}^s|c_j|: \hat{x}[l] = \sum\limits_{j=1}^{s} c_je^{i2\pi f_jl} \phantom{1}, \phantom{1} l \in \mathcal{M}\right\}
\end{equation}
To estimate the remaining $\mathcal{N} \setminus \mathcal{M}$ samples of the signal $x$, \cite{chandrasekaran2012theconvex} suggests minimizing the atomic norm $||\hat{x}||_\mathcal{A}$ among all vectors $\hat{x}$ leading to the same observed samples as $x$. Intuitively, the atomic norm minimization is similar to $\ell_1$-minimization being the tightest convex relaxation of the combinatorial $\ell_0$-minimization problem. The \textit{primal} convex optimization problem for atomic norm minimization can be formulated as follows,
\begin{flalign}
	\label{eq:atomicminimization}
	& \underset{\hat{x}}{\text{minimize}}\phantom{1}  \|\hat{x}\|_{\mathcal{A}}\nonumber\\
	& \text{subject to}\phantom{1} \hat{x}[j] = x[j], \nonumber\\
	& \phantom{1}\phantom{1}\phantom{1}\phantom{1}\phantom{1}\phantom{1}\phantom{1}\phantom{1}\phantom{1}\phantom{1} j \in T, \phantom{1} T = \{0, \cdots, n-1\}
\end{flalign}
Equivalently, the off-the-grid compressed sensing \cite{tang2012csotg} suggests the following semidefinite characterization for $||\hat{x}||_\mathcal{A}$:
\begin{definition} \cite{tang2012csotg} Let $T_n$ denote the $n \times n$ positive semidefinite Toeplitz matrix, $t \in \mathbb{R}^+$, Tr($\cdot$) denote the trace operator and $(\cdot)^*$  denote the complex conjugate. Then,
    \begin{equation}
	||\hat{x}||_{\mathcal{A}} = \underset{T_n, t}{\text{inf}} \left\{\dfrac{1}{2|\mathcal{N}|} \text{Tr($T_n$)} + \frac{1}{2}t : \begin{bmatrix*}[r] T_n & \hat{x} \\ \hat{x}^* & t \end{bmatrix*} \succeq 0 \right\}
	\end{equation}
\end{definition}

The positive semidefinite Toeplitz matrix $T_n$ is related to the frequency atoms through the following Vandemonde decomposition result by Carath{\`e}odory \cite{cara1911uber}:
\begin{align}
T_n &= URU^*\\
\text{where } U_{lj} &= a(f_{j}, \phi_{j})[l],\\
                    R &= \text{diag}([b_1, \cdots, b_{r}])
\end{align}
The diagonal elements of $R$ are real and positive, and $r = \text{rank}(T_n)$.

Consistent with this definition, the atomic norm minimization problem for the frequency-sparse signal recovery can now be formulated as a semidefinite program (SDP) with $m$ affine equality constraints:
\begin{flalign}
	\label{eq:semiotg}
	& \underset{T_n, \hat{x}, t}{\text{minimize}}\phantom{1} \dfrac{1}{2|\mathcal{N}|} \text{Tr($T_n$)} + \frac{1}{2}t\nonumber\\
	& \text{subject to}\phantom{1} \begin{bmatrix*}[r] T_n & \hat{x} \\ \hat{x}^* & t \end{bmatrix*} \succeq 0\\
	& \hat{x}[l] = x[l], \phantom{1} l \in \mathcal{M}\nonumber
\end{flalign}

\subsection{Frequency localization using dual norm}
\label{subsec:freq_local}
The frequencies in $\hat{x}$ can then be identified by the \textit{frequency localization} approach \cite{tang2012csotg} based on computing the dual-polynomial $Q_f^\star = \langle q^{\star}, a(f,0)\rangle$, where $(\cdot)^{\star}$ denotes quantities corresponding to the solution of the following dual problem of (\ref{eq:atomicminimization}):
\begin{flalign}
	\label{eq:dualtoatomicminimization}
	 \underset{q}{\text{maximize}} & \phantom{1} \langle q_{\mathcal{M}},x_{\mathcal{M}}\rangle_{\mathbb{R}}\phantom{1}  \nonumber\\
	 \text{subject to}&\phantom{1}\|q\|_{\mathcal{A}}^* \leq 1 \\
	&\phantom{1} q_{\mathcal{N}\setminus \mathcal{M}}=0 \nonumber
\end{flalign}
where $\|\cdot\|^*$ represents the dual norm. This dual norm is defined as
\begin{align}
\label{eq:dualatomicnorm}
\|q\|_{\mathcal{A}}^*=\sup_{\|\hat{x}\|_{\mathcal{A}}\leq 1} \langle q,\hat{x} \rangle_{\mathbb{R}}=\sup_{f \in [0,1]}|\langle q, a(f,0)\rangle|
\end{align}
For the frequency localization, $|Q_{f_j}| = 1$, if $f_j$ is one of the unknown frequencies of $x$. Otherwise, $|Q_{f_j}| < 1$.

\subsection{Using prior information}
\label{subsubsec:otgcs_prior_info}
A common approach to harness the prior information of the sparse signal in compressed sensing algorithms is to replace the classical $\ell_1$ norm with the weighted $\ell_1$ norm \cite{amin2009weighted} \cite{vaswani2010modified}. However, signals with continuous-valued frequencies do not lead to a trivial application of the weighted $\ell_1$ approach. When the frequencies are known to reside in a known set $C$ \textit{a priori}, we can minimize a \textit{constrained} atomic norm $||\hat{x}||_{\mathcal{A}, \mathcal{C}}$ \cite{mishra2013off}:
\begin{equation}
||\hat{x}||_{\mathcal{A}, \mathcal{C}} = \underset{c_j, f_j \in \mathcal{C}}{\text{inf}}\phantom{1}\left\{\sum\limits_{j=1}^s|c_j|: \hat{x}[l] = \sum\limits_{j=1}^{s} c_je^{i2\pi f_j l} \phantom{1}, \phantom{1} l \in \mathcal{M}\right\}
\end{equation}
where $\mathcal{C}$ is a known set of frequencies. The dual problem of minimizing the conditional atomic norm is similar to its analog in (\ref{eq:dualtoatomicminimization}):
\begin{flalign}
  \underset{q}{\text{maximize}} & \phantom{1} \langle q_{\mathcal{M}},x_{\mathcal{M}}\rangle_{\mathbb{R}}\phantom{1}  \nonumber\\
  \text{subject to}&\phantom{1}\|q\|_{\mathcal{A},\mathcal{C}}^* \leq 1\\
 &\phantom{1} q_{\mathcal{N}\setminus \mathcal{M}}=0 \nonumber
\end{flalign}
While the Vandemonde decomposition holds for general positive semidefinite Toeplitz matrices, it is not clear how to further tighten the Toeplitz structure to reflect the known prior information. Thus for an arbitrary set $\mathcal{C}$, formulating a computable convex program is not trivial for the constrained atomic norm minimization. In this paper, we propose a \textit{precise} semidefinite program for minimizing the conditional atomic norm when $\mathcal{C}$ corresponds to block priors.

\subsection{System model with block priors}
\label{subsec:anormmin_prior}
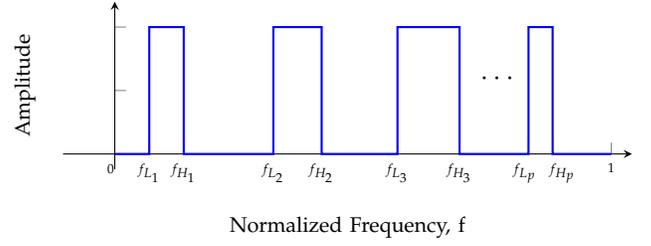
\begin{figure}
\centering
	\begin{tikzpicture}
	\begin{axis}[
		width=3.6in,
		height=1.5in,
		x axis line style={-stealth},
		y axis line style={-stealth},
		xtick={0.5, 1.0, 2.3, 3, 4.1, 5, 6.0, 6.35, 7.2},
		xticklabels={\tiny$f_{L_1}$, \tiny$f_{H_1}$, \tiny$f_{L_2}$, \tiny$f_{H_2}$, \tiny$f_{L_3}$, \tiny$f_{H_3}$, \tiny$f_{L_p}\phantom{1}$, \tiny$\phantom{1}\phantom{1}\phantom{1}f_{H_p}$, \tiny1},
		yticklabels={},
		ymax = 1.2,xmax=7.5,
		axis lines*=center,
		ylabel={\footnotesize Amplitude},
		xlabel={\footnotesize Normalized Frequency, f},
		xlabel near ticks,
		ylabel near ticks]
	\addplot+[thick,mark=none,const plot]
	coordinates
	{(0,0) (0.5,1) (1.0,0) (2.3,1) (3,0) (4.1,1) (5,0) (6.0,1) (6.35,0) (7.2,0)};
	\end{axis}	
	\node [left] at (0.8,  0) {\tiny 0};
	\node [above] at (5.8,  1.0) {$\cdots$};
	\end{tikzpicture}
	\caption{\small The individual frequencies of spectrally parsimonious signal are assumed to lie in known frequency subbands within the normalized frequency domain $[0, 1]$. While the system model doesn't impose an upper limit on $p$, it does assume that all subbands are non-overlapping.}
	\label{fig:mult_bands}
\end{figure}
Let us now consider the case where all the $s$ frequencies $f_j$ of the spectrally sparse signal $x$ are known \textit{a priori} to lie only in a finite number of non-overlapping frequency bands or intervals within the normalized frequency domain $[0, 1]$. Here, the known set $\mathcal{C}$ is defined as the set $\mathcal{B}$ of all the known frequency bands. The prior information consists of the precise locations of all the frequency bands - the lower and upper cut-off frequencies $f_{L_k}$ and $f_{H_k}$ respectively for each of the band $\mathcal{B}_k$ - as shown in the Figure \ref{fig:mult_bands}. We, therefore, have
\begin{align}
f_j \in \mathcal{B}, \phantom{1}  \mathcal{B} = \bigcup_{k = 1}^{p} \mathcal{B}_k = \bigcup_{k = 1}^{p} [f_{L_k}, f_{H_k}],
\end{align}
where $p$ is the total number of disjoint bands known \textit{a priori}.

\section{Semidefinite Program Formulation}
\label{sec:sdp}
As noted earlier, to recover all of the off-the-grid frequencies of the signal $x$ given the block priors, the direct extension of a semidefinite program from (\ref{eq:semiotg}) to minimize the constrained atomic norm is non-trivial. We address this problem by working with the dual problem of the constrained atomic norm minimization, and then transforming the dual problem to an equivalent semidefinite program by using theories of positive trigonometric polynomials. 

We note that in the case of block priors, (\ref{eq:dualatomicnorm}) can be written as
\begin{align}
\|q\|_{\mathcal{A}, \mathcal{B}}^* = \sup_{f \in \mathcal{B}}|\langle q, a(f,0)\rangle| = \sup_{f \in \mathcal{B}}|Q_f|
\end{align}
where $Q_f$ is the dual polynomial.

Here, the primal atomic norm minimization problem is given by
\begin{flalign}
	\label{eq:blockpriorprimal}
	& \underset{x}{\text{minimize}}\phantom{1}  \|x\|_{\mathcal{A}, \mathcal{B}}\nonumber\\
	& \text{subject to}\phantom{1} \hat{x}[l] = x[l], \phantom{1} l \in \mathcal{M}
\end{flalign}
Similar to (\ref{eq:dualtoatomicminimization}), we can formulate the corresponding dual problem as
\begin{flalign}
	\label{eq:dual_withblockpriors}
	 \underset{q}{\text{maximize}} & \phantom{1} \langle q_{\mathcal{M}},x_{\mathcal{M}}\rangle_{\mathbb{R}}\phantom{1}  \nonumber\\
	 \text{subject to}&\phantom{1} q_{\mathcal{N}\setminus \mathcal{M}}=0\\
	 &\phantom{1} \|q\|_{\mathcal{A}, \mathcal{B}}^* \leq 1 \nonumber
\end{flalign}
or equivalently,
\begin{flalign}
	\label{eq:dual_substituted}
	 \underset{q}{\text{maximize}} & \phantom{1} \langle q_{\mathcal{M}},x_{\mathcal{M}}\rangle_{\mathbb{R}}\phantom{1}  \nonumber\\
	 \text{subject to}&\phantom{1} q_{\mathcal{N}\setminus \mathcal{M}}=0\\
	 &\phantom{1} \sup_{f \in \mathcal{B}}|\langle q, a(f,0)\rangle| \leq 1 \nonumber
\end{flalign}
Since $\mathcal{B}$ is defined as a union of multiple frequency bands, the inequality constraint in (\ref{eq:dual_substituted}) can be expanded to $p$ separate inequality constraints as follows:
\begin{flalign}
	\label{eq:blockdual}
	 \underset{q}{\text{maximize}} & \phantom{1} \langle q_{\mathcal{M}},x_{\mathcal{M}}\rangle_{\mathbb{R}}\phantom{1}  \nonumber\\
	 \text{subject to}&\phantom{1} q_{\mathcal{N}\setminus \mathcal{M}}=0\\
	 &\phantom{1} \sup_{f \in [f_{L_1}, f_{H_1}]}|\langle q, a(f,0)\rangle| \leq 1 \nonumber\\
	 &\phantom{1} \sup_{f \in [f_{L_2}, f_{H_2}]}|\langle q, a(f,0)\rangle| \leq 1 \nonumber\\
	 & \phantom{1}\phantom{1}\phantom{1}\phantom{1}\phantom{1}\phantom{1}\phantom{1}\phantom{1}\phantom{1}\vdots \nonumber\\
	 &\phantom{1} \sup_{f \in [f_{L_p}, f_{H_p}]}|\langle q, a(f,0)\rangle| \leq 1 \nonumber
\end{flalign}
Our objective is to change each of the inequality constraints in (\ref{eq:blockdual}) to linear matrix inequalities, so that semidefinite programming is applicable to this problem.
\subsection{Gram matrix parametrization of positive trigonometric polynomials}
\label{subsec:gram_ptp}
We observe that $Q_f = \langle q, a(f,0)\rangle$ is a positive trigonometric polynomial in $f$ since
\begin{align}
\label{eq:dualpoly_expression}
Q_f = \langle q, a(f,0) \rangle = \sum\limits_{l=0}^{n-1} q_{l}e^{-i 2\pi f l }
\end{align}
and, as remarked in Section \ref{subsec:freq_local}, $|Q_f| \le 1$ for every $f \in \mathcal{B}$. A trigonometric polynomial, which is also nonnegative on the unit circle, can be parametrized using a positive semidefinite, \textit{Gram} matrix $\bm{Q}$ which allows description of such a polynomial using a linear matrix inequality \cite{dumitrescu2007positive}. For the trigonometric polynomial that is nonnegative only over a subinterval, we have the following theorem:
\begin{theorem} \cite[p. 12]{dumitrescu2007positive} A trigonometric polynomial
\begin{align}
R(z) &= \sum\limits_{k=-(n-1)}^{n-1}r_k z^{-k},\phantom{1} r_{-k} = r_k^*
\end{align}
where $R \in \mathbb{C}_n[z]$ for which $R(\omega) \ge 0$, for any $z = e^{i\omega}$, $\omega \in [\omega_L, \omega_H] \subset [-\pi, \pi]$, can be expressed as
\begin{align}
R(z) = F(z)F^*(z^{-1}) + D_{\omega_L\omega_H}(z).G(z)G^*(z^{-1})
\end{align}
where $F$, $G$ are causal polynomials with complex coefficients, of degree at most $n-1$ and $n-2$, respectively. The polynomial
\begin{align}
D_{\omega_L\omega_H}(z) &= d_1z^{-1} + d_0 + d_1^*z\\
\text{where } d_0 &= -\dfrac{\alpha\beta+1}{2}\\
d_1 &= \dfrac{1-\alpha\beta}{4} + j\dfrac{\alpha+\beta}{4}\\
\alpha &= \tan{\dfrac{\omega_L}{2}}\\
\beta &= \tan{\dfrac{\omega_H}{2}}
\end{align}
is defined such that $D_{\omega_L\omega_H}(\omega)$ is nonnegative for $\omega \in [\omega_L, \omega_H]$ and negative on its complementary.
\label{thm:dumitrescu_1_15}
\end{theorem}
Since $F$ and $G$ are causal polynomials, they can each be parameterized with Gram matrices $\bm{Q}_1$ and $\bm{Q}_2$ respectively, where $\bm{Q}_1 \in \mathbb{C}^{n \times n}$ and $\bm{Q}_2 \in \mathbb{C}^{(n-1) \times (n-1)}$ \cite[p. 23]{dumitrescu2007positive}. The polynomial $R$ can be parameterized using both of these Gram matrices as follows:
\begin{align}
\label{eq:lmi_freq}
r_k &= tr[\mathbf{\Theta}_k \bm{Q}_1] + \Tr{[(d_1\mathbf{\Theta}_{k-1} + d_0\mathbf{\Theta}_{k} + d_1^*\mathbf{\Theta}_{k+1}) \cdot \bm{Q}_2]}\nonumber\\
	&\triangleq \mathcal{L}_{k, \omega_L, \omega_H}(\bm{Q}_1, \bm{Q}_2)
\end{align}
where $\mathbf{\Theta}_k$ is the elementary Toeplitz matrix with ones on the k-th diagonal and zeros elsewhere. In the argument of the second trace operator, we assume $\mathbf{\Theta}_k = 0$ if $k > n-2$.

If, instead of $[\omega_L, \omega_H] \subset [-\pi, \pi]$, the subbands are expressed in the normalized frequency domain as $[f_L, f_H] \subset [0, 1]$, then (\ref{eq:lmi_freq}) can be written as,
\begin{align}
\label{eq:lmi_normfreq}
r_k &\triangleq \mathcal{L}_{k, f_L, f_H}(\bm{Q}_1, \bm{Q}_2)
\end{align}
where the translation of frequencies between the two domains is given by these relations:
\begin{align}
f_L &= \begin{dcases*}
        			2\pi\omega_L & : 0 $\le$ $\omega_L$ $\le$ 0.5\\
        			2\pi(\omega_L - 1) & : 0.5 < $\omega_L$ $\le$ 1
        	   \end{dcases*}\\
f_H &= \begin{dcases*}
        			2\pi\omega_H & : 0 $\le$ $\omega_H$ $\le$ 0.5\\
        			2\pi(\omega_H - 1) & : 0.5 < $\omega_H$ $\le$ 1
        	 \end{dcases*}
\end{align}
The dual polynomial $Q_f$ in (\ref{eq:dualpoly_expression}) is nonnegative on multiple non-overlapping intervals, and can therefore be parameterized by $p$ different pairs of Gram matrices $\{\bm{Q}_1$, $\bm{Q}_2\}$.
\subsection{LMI representation}
\label{subsec:lmi_ptp}
Based on the Bounded Real Lemma \cite[p. 127]{dumitrescu2007positive} (which, in turn, is based on Theorem \ref{thm:dumitrescu_1_15}), a positive trigonometric polynomial constraint of the type $|R(\omega)| \leq 1$ can be expressed as a linear matrix inequality \cite[p. 143]{dumitrescu2007positive}. Stating this result for the dual polynomial constraint over a single frequency band (such as those in (\ref{eq:blockdual})), we have $\sup_{f \in [f_L, f_H]}|\langle q, a(f,0)\rangle| \leq 1$, if and only if there exist positive semidefinite matrices $\bm{Q}_1$ and $\bm{Q}_2$ such that,
\begin{align}
\label{eq:lmi_block}
\delta_k = \mathcal{L}_{k, f_L, f_H} (\bm{Q}_1, \bm{Q}_2), & \phantom{1} k \in \mathcal{H} \nonumber\\
\begin{bmatrix} \bm{Q}_1 & q \\ q^{*} & 1 \end{bmatrix} &\succeq 0
\end{align}
where $\delta_0 = 1$ and $\delta_k = 0$ if $k \neq 0$ and $\mathcal{H}$ is a halfspace. This linear matrix inequality representation using positive semidefinite matrix $\bm{Q}_1$ paves way for casting the new dual problem in (\ref{eq:blockdual}) as a semidefinite program.

\subsection{Continuous compressed sensing with block priors}
\label{subsec:bp_ccs}
We are now in a position to state our semidefinite program for continuous compressed sensing with block priors. For each of the inequality constraint in (\ref{eq:blockdual}), we use a linear matrix inequality similar to that in (\ref{eq:lmi_block}) to cast the dual problem constraint into a semidefinite program. So, when all the frequencies are known to lie in $p$ disjoint frequency bands, then semidefinite program for the dual problem in (\ref{eq:blockdual}) can be constructed by using $p$ constraints of the kind in (\ref{eq:lmi_block}):
\fbox{
 \addtolength{\linewidth}{-2\fboxsep}%
 \addtolength{\linewidth}{-2\fboxrule}%
 \begin{minipage}{\linewidth}
\begin{flalign}
	\label{eq:blocksparsitySDP}
	\underset{\begin{subarray}{c}
	 q,\\
	 \bm{Q}_{11}, \bm{Q}_{12}, \cdots, \bm{Q}_{1p},\\
	 \bm{Q}_{21}, \bm{Q}_{22}, \cdots, \bm{Q}_{2p}
	\end{subarray}}{\text{maximize}} &\phantom{1} \langle q_{\mathcal{M}},x_{\mathcal{M}} \rangle_{\mathbb{R}}\phantom{1}  \nonumber\\
	\text{subject to   }&\phantom{1} q_{\mathcal{N}\setminus \mathcal{M}}=0\phantom{1}\phantom{1}\phantom{1}\phantom{1}\phantom{1}\phantom{1}\phantom{1}\\
	&\phantom{1} \delta_k = \mathcal{L}_{k, f_{L_1}, f_{H_1}} (\bm{Q}_{11}, \bm{Q}_{21}),\nonumber\\
	& \phantom{1}\phantom{1}\phantom{1}\phantom{1}\phantom{1}\phantom{1}\phantom{1} k = 0, \cdots, (n-1)\nonumber\\
	& \begin{bmatrix*}[r] \bm{Q}_{11} & q \\ q^* & 1 \end{bmatrix*} \succeq 0,\nonumber\\
	&\phantom{1} \delta_k = \mathcal{L}_{k, f_{L_2}, f_{H_2}} (\bm{Q}_{12}, \bm{Q}_{22}),\nonumber\\
	& \phantom{1}\phantom{1}\phantom{1}\phantom{1}\phantom{1}\phantom{1}\phantom{1} k = 0, \cdots, (n-1)\nonumber\\
	& \begin{bmatrix*}[r] \bm{Q}_{12} & q \\ q^* & 1 \end{bmatrix*} \succeq 0,\nonumber\\
	&\phantom{1}\phantom{1}\phantom{1}\phantom{1}\phantom{1}\phantom{1}\phantom{1}\phantom{1}\phantom{1}\phantom{1}\phantom{1}\phantom{1}\vdots\nonumber\\
	&\phantom{1} \delta_k = \mathcal{L}_{k, f_{L_p}, f_{H_p}} (\bm{Q}_{1p}, \bm{Q}_{2p}),\nonumber\\
	& \phantom{1}\phantom{1}\phantom{1}\phantom{1}\phantom{1}\phantom{1}\phantom{1} k = 0, \cdots, (n-1)\nonumber\\
	& \begin{bmatrix*}[r] \bm{Q}_{1p} & q \\ q^* & 1 \end{bmatrix*} \succeq 0,\nonumber\\
	\text{where }&\bm{Q}_{11}, \bm{Q}_{12}, \cdots, \bm{Q}_{1p} \in \mathbb{C}^{n \times n},\nonumber\\
	\text{and }&\bm{Q}_{21}, \bm{Q}_{22}, \cdots, \bm{Q}_{2p} \in \mathbb{C}^{(n-1) \times (n-1)}\nonumber
\end{flalign}
\end{minipage}
}

\section{Numerical Experiments}
\label{sec:numsim}
\begin{figure}[!t]
	\centerline{\subfloat[Without any priors]{\includegraphics[width=3in]{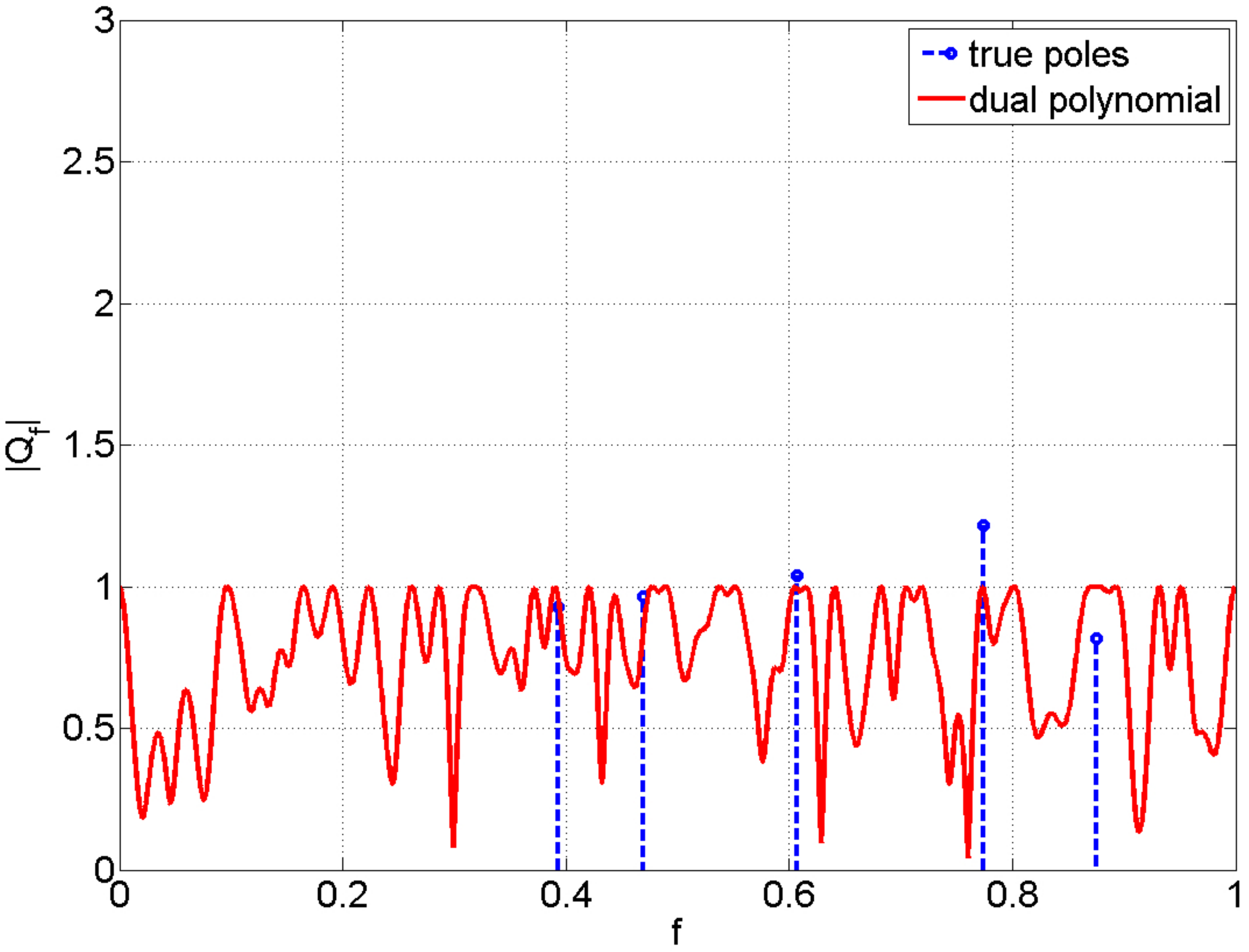}
	\label{fig:no_priors}}}
	\hfil	
\subfloat[With block priors]{\includegraphics[width=3in]{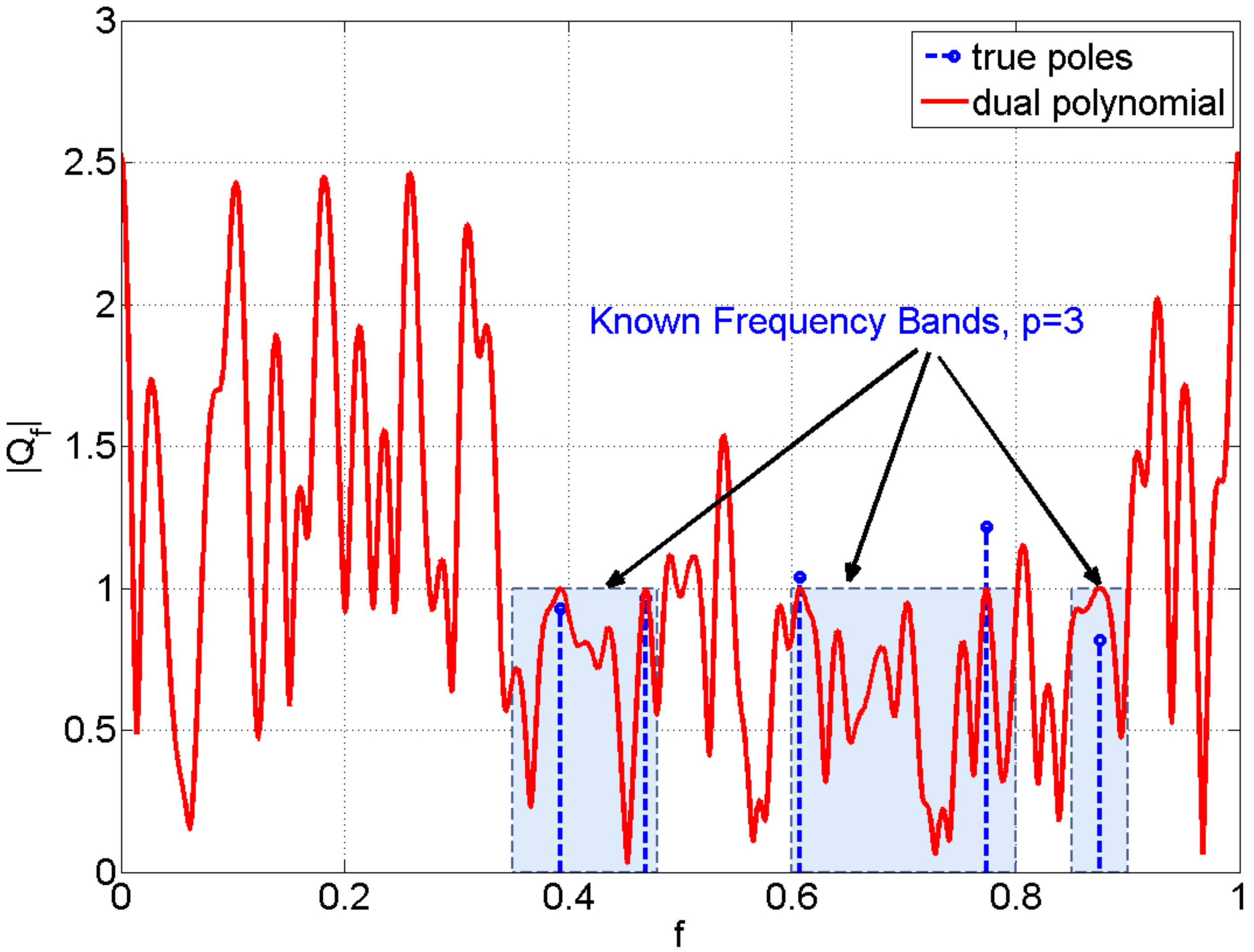}
\label{fig:with_priors}}
\caption{\small Frequency localization using dual polynomial for $\{n, s, m\} = \{64, 5, 20\}$. The block priors are $\mathcal{B} =$ $(0.3500, 0.4800)$ $\bigcup$ $(0.6000, 0.8000)$ $\bigcup$ $(0.8500, 0.9000)$.}
\label{fig:recovery_illust}
\end{figure}

We evaluated the performance of spectrum estimation with block priors through numerical simulations. We used SDPT3 \cite{tutuncu2003solving} solver for the semidefinite program in (\ref{eq:blocksparsitySDP}).

We first illustrate our approach through an example in  Figure \ref{fig:recovery_illust}. Here for $n = 64$, we drew $s = 5$ frequencies uniformly at random within $p = 3$ subbands in the domain $[0, 1]$ without imposing any minimum separation condition. Here, $\mathcal{B} =$ $(0.3500, 0.4800)$ $\bigcup$ $(0.6000, 0.8000)$ $\bigcup$ $(0.8500, 0.9000)$. The phases of the signal frequencies were sampled uniformly at random in $[0, 2\pi)$. The amplitudes $|c_j|, j = 1, \cdots, s$ were drawn randomly from the distribution $0.5 + \chi^2_1$ where $\chi^2_1$ represents the Chi-squared distribution with 1 degree of freedom. A total of $m = 20$ observations were randomly chosen from $n$ regular time samples to form the sample set $\mathcal{M}$. In the absence of any prior information, we solve (\ref{eq:dualtoatomicminimization}) and show the result of frequency localization in Figure \ref{fig:no_priors}. Here, it is difficult to pick a unique set of $s = 5$ poles for which the maximum modulus of the dual polynomial is unity (which will actually correspond to recovered frequency poles). On the other hand, when block priors are given, Figure \ref{fig:with_priors} shows that solving (\ref{eq:blocksparsitySDP}) provides perfect recovery of all the frequency components, where the recovered frequencies correspond to unit-modulus points of the dual polynomial.

\begin{figure}[!t]
\centering
\includegraphics[width=3.5in]{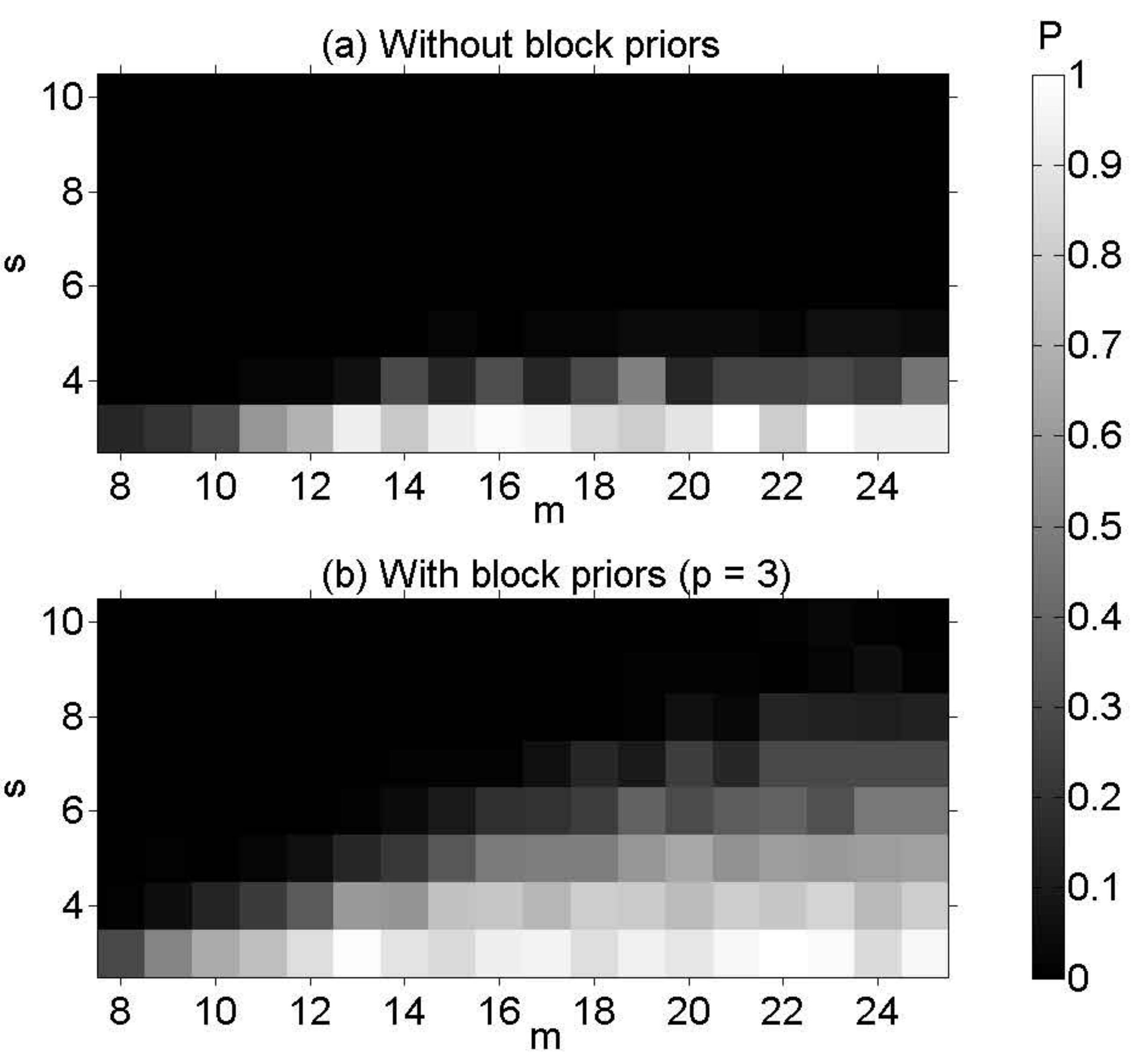}
\caption{\small The probability $P$ of perfect recovery over 100 trials for $n = 64$. The block priors are $\mathcal{B} =$ $(0.3500, 0.4800)$ $\bigcup$ $(0.6000, 0.8000)$ $\bigcup$ $(0.8500, 0.9000)$. For each realization of the signal, the $s$ frequencies were chosen uniformly at random within the interval $\mathcal{B}$ without imposing any minimum frequency spacing.}
\label{fig:statcomp}
\end{figure}

We then give a statistical performance evaluation of our new method, compared with atomic norm minimization without any priors (\ref{eq:dualtoatomicminimization}). The experimental setup and block priors are the same as in Figure \ref{fig:recovery_illust} and no minimum separation condition was assumed while drawing frequencies uniformly at random in the set $\mathcal{B}$. Figure \ref{fig:statcomp} shows the probability $P$ of perfect recovery for the two methods for fixed $n=64$ but varying values of $m$ and $s$. For every value of the pair $\{s, m\}$, we simulate 100 trials to compute $P$. We note that if the frequencies are approximately known, our method greatly enhances the recovery of continuous-valued frequencies.

\bibliographystyle{IEEEtran}
\bibliography{refs}

\end{document}